\documentclass[12pt,fleqn]{article}
\pdfoutput=1
\usepackage{amsmath}
\usepackage{amsthm}
\usepackage{amsfonts}
\usepackage{amssymb}
\usepackage{amscd}
\usepackage{fullpage}
\usepackage{color}
\usepackage{times}
\usepackage{graphicx}
\usepackage{float}
\usepackage{hyperref}

\begin{document}
\renewcommand{\thefootnote}{\fnsymbol{footnote}}
\title{On Backus average for oblique incidence}
\author{
David R. Dalton\footnote{
Department of Earth Sciences, Memorial University of Newfoundland,
{\tt n06drd@mun.ca}},
Ayiaz Kaderali\footnote{
Department of Earth Sciences, Memorial University of Newfoundland,\newline
{\tt ayiazkaderali@gmail.com}}
}
\date{\today}
\maketitle
\renewcommand{\thefootnote}{\arabic{footnote}}
\setcounter{footnote}{0}
\section*{Abstract}
We postulate that validity of the Backus (1962) average, whose weights are layer thicknesses, is limited to waves whose incidence is nearly vertical.
The accuracy of this average decreases with the increase of the source-receiver offset.
However, if the weighting is adjusted by the distance travelled by a signal in each layer, such a modified average results in a more accurate prediction of traveltimes through these layers.
\section{Introduction}
Hookean solids, which are commonly used in seismology as mathematical analogies of physical materials, are defined by their mechanical property relating linearly the stress tensor,~$\sigma$\,, and the strain tensor,~$\varepsilon$\,,
\begin{equation*}
\sigma_{ij}=\sum_{k=1}^3\sum_{\ell=1}^3c_{ijk\ell}\varepsilon_{k\ell}\,,\qquad i,j=1,2,3
\,,
\end{equation*}
where $c$ is the elasticity tensor.
The Backus (1962) average allows us to quantify the response of a wave propagating through a series of parallel Hookean layers whose thicknesses are much smaller than the wavelength.

According to Backus (1962), the average of $f(x_3)$ of ``width''~$\ell'$  is
\begin{equation}
\label{eq:BackusOne}
\overline f(x_3):=\int\limits_{-\infty}^\infty w(\zeta-x_3)f(\zeta)\,{\rm d}\zeta
\,,
\end{equation}
where $w(x_3)$ is the weight function with the following properties:
\begin{eqnarray*}
&&w(x_3)\geqslant0\,,
\quad w(\pm\infty)=0\,,
\quad
\int\limits_{-\infty}^\infty w(x_3)\,{\rm d}x_3=1\,,\\
&&\int\limits_{-\infty}^\infty x_3w(x_3)\,{\rm d}x_3=0\,,
\quad
\int\limits_{-\infty}^\infty x_3^2w(x_3)\,{\rm d}x_3=(\ell')^2\,.
\end{eqnarray*}
These properties define $w(x_3)$ as a probability-density function with mean~$0$ and standard deviation~$\ell'$\,, explaining the use of the term ``width'' for $\ell'$\,.

The long-wavelength homogeneous media equivalent to a stack of isotropic or transversely isotropic layers with thicknesses much less than the signal wavelength are shown by Backus (1962) to be transversely isotropic.
The Backus (1962) formulation is reviewed by Slawinski (2018) and Bos et al.\ (2017),
where formulations for generally anisotropic, monoclinic, and orthotropic thin layers
are also derived.
Bos et al.\ (2017) examine assumptions and approximations underlying the Backus (1962) formulation, which is derived by expressing rapidly varying stresses and strains in terms of products of algebraic combinations of rapidly varying elasticity parameters with slowly varying stresses and strains.
The only mathematical approximation in the formulation
is that the average of a product of a rapidly varying function and a slowly
varying function is approximately equal to the product of the averages of
the two functions.

From Slawinski (2018), if isotropic layers are described by the elasticity parameters $c_{1111}$ and $c_{2323}$\,, the corresponding parameters of the transversely isotropic medium are
\begin{equation}
\label{eq:Tue1}
c^{\overline{\rm
TI}}_{1111}=\overline{\left(\dfrac{c_{1111}-2c_{2323}}{c_{1111}}\right)}^{\,2}
\,\,\,\overline{\left(\dfrac{1}{c_{1111}}\right)}^{\,-1}
+\overline{\left(\dfrac{4(c_{1111}-c_{2323})c_{2323}}{c_{1111}}\right)}\,,
\end{equation}
\begin{equation}
\label{eq:Mon10}
c^{\overline{\rm
TI}}_{1122}=\overline{\left(\dfrac{c_{1111}-2c_{2323}}{c_{1111}}\right)}^{\,2}
\,\overline{\left(\dfrac{1}{c_{1111}}\right)}^{\,-1}
+\overline{\left(\dfrac{2(c_{1111}-2c_{2323})c_{2323}}{c_{1111}}\right)}\,,
\end{equation}
\begin{equation}
\label{eq:Backus1133}
c^{\overline{\rm TI}}_{1133}=\overline{\left(\dfrac{c_{1111}-2c_{2323}}{c_{1111}}\right)}\,\,
\,\,\overline{\left(\dfrac{1}{c_{1111}}\right)}^{\,-1}
\,,
\end{equation}
\begin{equation}
\label{eq:Berry1}
c^{\overline{\rm TI}}_{1212}=\overline{c_{2323}}
\,,
\end{equation}
\begin{equation}
\label{eq:Berry2}
c^{\overline{\rm TI}}_{2323}=\overline{\left(\dfrac{1}{c_{2323}}\right)}^{\,-1}
\,,
\end{equation}
\begin{equation}
\label{eq:Tue2}
c^{\overline{\rm TI}}_{3333}=\overline{\left(\dfrac{1}{c_{1111}}\right)}^{\,-1}
\,,
\end{equation}
which are the {\sl Backus parameters} for isotropic layers.
\section{Formulation for a ten-layer synthetic model}
Let us consider a stack of ten isotropic horizontal layers, each with a thickness of $100$ meters (Brisco, 2014).
Their elasticity parameters are listed in \ref{app:tenlayer}, Table~\ref{table:BackusAverage}.

For vertical incidence, the Fermat traveltime through these layers is $229.46$~ms.
If we perform the standard Backus average---weighted by layer thickness of these ten layers, as in equation~(\ref{eq:WeightEG}), below, which is derived in \ref{app:backus-arith}---then, the equivalent 
density-scaled elasticity parameters, in units of $10^6~{\rm m}^2/{\rm s}^2$\,, are
$\langle c_{1111}\rangle =18.84$\,, $\langle c_{1212}\rangle =3.99$\,, $\langle c_{1133}\rangle =10.96$\,, $\langle c_{2323}\rangle =3.38$ and $\langle c_{3333}\rangle =18.43$\,.
With these parameters and for the vertical incidence, the resulting $P$-wave traveltime through the equivalent transversely isotropic medium is $232.91$~ms, which---in comparison to the Fermat traveltime---is high by $3.45$~ms.    
We also tried a weighting by the traveltime in each layer, and that resulted in, for vertical incidence, the resulting $P$-wave traveltime through the equivalent transversely isotropic medium of $239.76$~ms, which---in comparison to the Fermat traveltime---is high by $10.30$~ms.  
Thus we did not consider weighting by traveltime further.

To examine the layer-thickness weighting, let us consider one of the equivalent-medium parameters,
\begin{equation}
\label{eq:WeightEG}
c^{\overline{\rm TI}}_{1212}=\frac{\sum\limits_{i=1}^nh_i\,{c_{2323}}_i}{\sum\limits_{j=1}^nh_j}
 \,,
\end{equation}
where $h_i$ is the thickness of the $i$th layer, which herein is $100$~m for each layer; thus, each layer is weighted equally by $0.1$\,.
\subsection{Fixed takeoff angle of \texorpdfstring{$\pi/6$}{pi/6}}
If we consider a $P$-wave signal whose takeoff angle, with respect to the vertical, is $\pi/6$\,, this signal reaches---in accordance with Snell's law---the bottom of the stack at a horizontal distance of $1072.53$~m.
Its Fermat traveltime is $330.52$~ms.

If we perform the standard Backus average, the traveltime in the equivalent medium, which corresponds to the ray angle of $47.01^\circ$\,, is $343.82$~ms, which is higher by $13.3$~ms than its Fermat counterpart.   
This is obtained using the equations in \ref{app:rayvelangle}.
If, however, we weight the average by the distance travelled in each layer, as in equation~(\ref{eq:WeightSlant}), below, the equivalent elasticity parameters become
$\langle c_{1111}\rangle =20.13$\,, $\langle c_{1212}\rangle =4.10$\,, $\langle c_{1133}\rangle =12.06$\,, $\langle c_{2323}\rangle =3.45$ and $\langle c_{3333}\rangle =19.76$\,.
In such a case, the traveltime is $332.36$~ms---which is higher by only $1.9$~ms---and is an order of magnitude more accurate than using the standard approach.   
The distances travelled in each layer and the resulting weights are given in \ref{app:tenlayer}, Table~\ref{table:weightings}, columns~2 and~3.
In such a case, expression~(\ref{eq:WeightEG}) becomes
\begin{equation}
\label{eq:WeightSlant}
c^{\overline{\rm TI}}_{1212}=\frac{\sum\limits_{i=1}^nd_i\,{c_{2323}}_i}{\sum\limits_{j=1}^nd_j}
 \,,
\end{equation}
where $d_i$ is the distance travelled in the $i$th layer, which---for vertical incidence---is equal to $h_i$\,.

According to Lemma~2 of Bos et al. (2017), the stability conditions are preserved by the Backus average.
In other words, if the individual layers satisfy these conditions, so does their equivalent medium.
This remains true for the modified Backus average.
\subsection{Extreme oblique example}
\label{sec:extremeoblique}
Let us now consider an extreme oblique example using the same ten layer model  as in the previous section but doing two-point ray-tracing with an offset of $7000~{\rm m}$ which results in a takeoff angle of $0.61026$ radians and a ray angle in the equivalent medium of $1.4289$ radians, which is $81.87^\circ$.

Given that, the Fermat traveltime is $1365.0~{\rm ms}$, and the thickness-weighted Backus average medium has the same elasticity parameters as in the previous section and a corresponding traveltime of $1343.1~{\rm ms}$, which is too high by $266.3~{\rm ms}$.    
The distances travelled in each layer and the resulting weights are given in \ref{app:tenlayer}, Table~\ref{table:weightings}, columns~4 and~5.
The slant-distance-weighted Backus average medium elasticity parameters become
$\langle c_{1111}\rangle =27.73$\,, $\langle c_{1212}\rangle =3.52$\,, $\langle c_{1133}\rangle =21.04$\,, $\langle c_{2323}\rangle =3.16$ and $\langle c_{3333}\rangle =26.08$\,.  
The corresponding traveltime, again obtained using the equations in \ref{app:rayvelangle}, is $1343.1~{\rm ms}$ which is too low by $21.0~{\rm ms}$.
Thus the slant-distance-weighting performs better than the thickness weighting again.
\section{A real data example}
{We now examine a real data example consisting of 15631 layers taken from a well log from a well offshore Newfoundland and extending over a height of $1595~{\rm m}$ from a depth of $1383~{\rm m}$ to a depth of $2978~{\rm m}$.    
Table~\ref{table:well-log} in \ref{app:well-log} gives layer thicknesses, $P$-wave velocities, and $S$-wave velocities for the first ten and last ten layers. 
The layer density-scaled elasticity parameters $c_{1111}$ and $c_{2323}$ are the squares of the  $P$-wave velocities and $S$-wave velocities, respectively.

For vertical incidence, the Fermat traveltime through these layers is $510.2$~ms.
If we perform the standard Backus average---weighted by layer thickness of these layers, as in equation~(\ref{eq:WeightEG}), the equivalent 
density-scaled elasticity parameters, in units of $10^6~{\rm m}^2/{\rm s}^2$\,, are
$\langle c_{1111}\rangle =10.75$\,, $\langle c_{1212}\rangle =3.29$\,, $\langle c_{1133}\rangle =4.10$\,, $\langle c_{2323}\rangle =2.42$ and $\langle c_{3333}\rangle =9.46$\,.
With these parameters and for the vertical incidence, the resulting $P$-wave traveltime through the equivalent transversely isotropic medium is $518.5$~ms, which---in comparison to the Fermat traveltime---is high by $8.3$~ms.

We consider a takeoff angle of $0.32$ radians, which is $18.3^\circ$.
If, we weight the average by the distance travelled in each layer, as in equation~(\ref{eq:WeightSlant}), the equivalent elasticity parameters become $\langle c_{1111}\rangle =11.05$\,, $\langle c_{1212}\rangle =3.41$\,, $\langle c_{1133}\rangle =4.14$\,, $\langle c_{2323}\rangle =2.49$ and $\langle c_{3333}\rangle =9.67$\,.

For a takeoff angle of $0.32$ radians, the Fermat traveltime is $581.2~{\rm ms}$, the thickness-weighted Backus average medium traveltime is $597.6~{\rm ms}$, which is too high by $16.4~{\rm ms}$, and the slant-weighted Backus average medium traveltime is $591.1~{\rm ms}$, which is too high by $9.9~{\rm ms}$.
Thus the slant-distance weighting performs better than the thickness weighting again though even the slant-distance weighting is still off by a fair bit, maybe because the Fermat calculation assumes high frequency and the Backus calculation assumes low frequency.
\section{Discussion}
The Backus (1962) average with weighting by the thickness of layer assumes vertical or near-vertical incidence.
Consequently, such an average does not result in accurate traveltimes for the far-offset or, in particular, cross-well data, which nowadays are common seismic experiments, and were not half-a-century ago, when the Backus (1962) average was formulated.

If we modify the weighting to be by the distance travelled in each layer, then the resulting traveltimes are significantly more accurate.
Such weighting, however, entails further considerations.
Since the distance travelled in each layer is a function of Snell's law, there is a need to modify the weights with the source-receiver offset.
However, given information about layers, it is achievable algorithmically by accounting for distance travelled in each layer as a function of offset.

There is also an interesting issue to consider. The modified equivalent medium is defined by its elasticity parameters, which are functions of the obliqueness of rays within each layer.
This means that the equivalent-medium parameters are different for the $qP$ waves, for the $qSV$ waves and for the $SH$ waves.
However, since a Hookean solid exists in the mathematical realm, not the physical world, such a consideration is not paradoxical.
It is common to invoke even different constitutive equations for the same physical material depending on empirical considerations.
Furthermore, it might be possible to derive elasticity parameters of a single Hookean solid---possibly of a material symmetry lower than transverse isotropy---whose behaviour accounts for both near and far offsets in the case of three waves.

It is interesting to note that, except for the slant-distance weighted equivalent medium for the extreme oblique model, in each case the traveltime in the equivalent medium is greater than its Fermat counterpart through the sequence of layers. It might be a consequence of optimization, which---in the case of layers---benefits from a model with a larger number of parameters.

There remains a fundamental question: Is the Fermat traveltime an appropriate criterion to consider the accuracy of the Backus average?
An objection to such a criterion is provided by the following {\it Gedankenexperiment\/}.
Consider a stack of thin layers, where---in one of these layers---waves propagate much faster than in all others.
In accordance with Fermat's principle, distance travelled by a signal within this layer is much larger than in any other layer, which might be expressed by the ratio of a distance travelled in a given layer divided by its thickness.
This effect is not accommodated by the standard Backus average, since this effect is offset-dependent and the average is not, but it is accommodated by the modified average discussed herein.
However, a property of such a single layer might be negligible on long-wavelength signal.
To address such issues, it might be necessary to consider a full-waveform forward model, and even a laboratory experimental set-up.

For the fastest layer, as for $v_P$ in layer five of Table~\ref{table:BackusAverage}, the distance travelled in that layer becomes much greater than the distance travelled in other layers as the ray angle with respect to the vertical in the fastest layer approaches $90^\circ$, and the takeoff angle approaches the maximum takeoff angle.
This is exemplified in columns 4 and 5 of Table~\ref{table:weightings} and in section~\ref{sec:extremeoblique}.

As an aside, let us recognize that---if we keep the Fermat traveltime as a criterion---making the propagation speed a function of the wavelength would not accommodate the traveltime discrepancy due to offset.

Be that as it may, it must be recognized that the discrepancy between the traveltimes in the layered and equivalent media increases with the source-receiver offset.
In the limit---for a wave propagating horizontally through a stack of horizontal layers---the Backus average, even in its modified form, is not valid, due to its underlying assumption of a load on the top and bottom only.
\section*{Acknowledgments}
We wish to acknowledge Python coding for verification of the takeoff angle for the $7000~{\rm m}$ two-point raytracing by Thomas B. Meehan, discussions with Michael A. Slawinski, and discussions with and Mathematica code for \ref{app:rayvelangle} from Theodore Stanoev. 
This research was performed in the context of The Geomechanics Project
supported by Husky Energy. Also, this research was partially supported by the
Natural Sciences and Engineering Research Council of Canada, grant 238416-2013.
\section*{References}
\frenchspacing
\newcommand{\hd}{\par\noindent\hangindent=0.4in\hangafter=1}
\hd
Backus, G.E.,  Long-wave elastic anisotropy produced by horizontal layering,
{\it  J. Geophys. Res.\/}, {\bf 67}, 11, 4427--4440, 1962.
\setlength{\parskip}{4pt}
\hd
Brisco, C., Anisotropy vs. inhomogeneity: Algorithm formulation, coding and modelling,
Honours Thesis, Memorial University, 2014.
\hd 
Bos, L, D.R. Dalton, M.A. Slawinski and T. Stanoev,
On Backus average for generally anisotropic layers, {\it Journal of Elasticity\/}, {\bf 127} (2), 179--196, 2017.
\hd
Slawinski, M.A. {\it Wavefronts and rays in seismology: Answers to unasked questions\/},
2nd edition, World Scientific, Singapore, 2018.
\hd
Slawinski, M.A., {\it Waves and rays in elastic continua\/}, 3rd edition, World Scientific, Singapore, 2015.
\setcounter{section}{0}
\setlength{\parskip}{0pt}
\renewcommand{\thesection}{Appendix~\Alph{section}}
\section{The ten-layer synthetic model}
\label{app:tenlayer}
\begin{table}[H]
\begin{center}
\begin{tabular}{|c||c|c|c|c|c|}
\hline
Layer & $c_{1111}$ & $c_{2323}$ & $v_P$ & $v_S$ \\
\hline\hline
1 & 10.56 & 2.02 & 3.25 & 1.42 \\
2 & 20.52 & 4.45 & 4.53 & 2.11 \\
3 & 31.14 & 2.89 & 5.58 & 1.70 \\
4 & 14.82 & 2.62 & 3.85 & 1.62 \\
5 & 32.15 & 2.92 & 5.67 & 1.71 \\
6 & 16.00 & 2.56 & 4.00 & 1.60 \\
7 & 16.40 & 6.35 & 4.05 & 2.52 \\
8 & 18.06 & 4.33 & 4.25 & 2.08 \\
9 & 31.47 & 8.01 & 5.61 & 2.83 \\
10 & 17.31 & 3.76 & 4.16 & 1.94 \\
\hline
\end{tabular}
\end{center}
\caption{\small{Density-scaled elasticity parameters, whose units are $10^6\,{\rm m}^{-2}{\rm s}^{-2}$\,, for a stack of isotropic layers, and the corresponding $P$-wave and $S$-wave speeds in ${\rm km\,s}^{-1}$\,.}}
\label{table:BackusAverage}
\end{table}
\begin{table}[H]
\begin{center}
\begin{tabular}{|c||c|c|c|c|}
\hline
layer & $d_i(\pi/6)$ & $w_i(\pi/6)$&$d_i(7000)$ & $w_i(7000)$\\
\hline\hline
1 & 115.47 & 0.0773&122.0&0.0167\\
2 & 139.45 & 0.0934&166.2&0.0228\\
3 & 195.07 & 0.1306&560.1&0.0767 \\
4 & 124.12 & 0.0831&136.2&0.0188\\
5 & 204.61 & 0.1370&5056.4&0.6921 \\
6 & 126.88 & 0.0849 &141.1&0.0193\\
7 & 127.85 & 0.0855&142.9 &0.0196\\
8 & 132.17 & 0.0885&151.0&0.0207\\
9 & 198.04 & 0.1326 &693.0&0.0935\\
10 & 130.17  & 0.0871&147.1&0.0201\\
\hline
\end{tabular}
\end{center}
\caption{Distances,~$d_i$\,, in meters,  travelled by the $P$ wave in each layer, and the corresponding averaging weights, $w_i=d_i/(\sum_{j=1}^{10} d_j)$\,, for a takeoff angle of $\pi/6$ and for an offset of $7000~{\rm m}$\,.}
\label{table:weightings}
\end{table}
\section{Thickness-weighted arithmetic average derivation}
\label{app:backus-arith}
\[
\overline{f}(x_3)=\int_{-\infty}^\infty w(\xi-x_3)f(\xi)\,{\rm d}\xi\,.
\]
Let
\[
w(y)=\frac{1}{2\sqrt{3}\ell'}{\rm I}_{[-\sqrt{3}\ell',\sqrt{3}\ell']}=
\begin{cases}
\frac{1}{2\sqrt{3}\ell'}&-\sqrt{3}\ell'\leq y\leq\sqrt{3}\ell'\\
0&y< -\sqrt{3}\ell'\mbox{ or }y>\sqrt{3}\ell'
\end{cases}
\]
Then if $Z=2\sqrt{3}\ell'$\,,
\[
w(y)=
\begin{cases}
1/Z&-Z/2\leq y\leq Z/2\\
0&y< -Z/2\mbox{ or }y>Z/2
\end{cases}
\]
Then if we let the midpoint be $x_3=Z/2=\sqrt{3}\ell'$\,,
\[
w(\xi-x_3)=w(\xi-Z/2)=
\begin{cases}
1/Z&0\leq \xi\leq Z\\
0&\xi<0\mbox{ or }\xi>Z
\end{cases}
\]
So
\[
\overline{f}(Z/2)=\frac{1}{Z}\int_0^Zf(\xi)\,{\rm d}\xi=
\frac{1}{Z}\sum_{i=1}^n h_i f_i\,,
\]
where $Z=\sum h_i$ is the total height,
and if $h_i$ is constant over all layers,
\[
\overline{f}(Z/2)=\frac{1}{n}\sum_{i=1}^n f_i\,.
\]
\section{The real well-log data}
\label{app:well-log}
\begin{table}[H]
\begin{center}
\begin{tabular}{|c||c|c|c|c|}
\hline
Layer & $h_i$ & $v_P$ & $v_S$ \\
\hline\hline
1 & 0.0971334 & 2131.23 & 1017.06 \\
2 & 0.0971334 & 2165.30 & 1019.65  \\
3 & 0.0971334 & 2230.32 & 1029.47  \\
4 & 0.0971334 & 2320.83 & 1039.11 \\
5 & 0.0971334 & 2409.92 & 1050.14  \\
6 & 0.0971334 & 2463.18 & 1067.63 \\
7 & 0.0971334 & 2496.51 & 1081.11  \\
8 & 0.0971334 & 2505.24 & 1088.57 \\
9 & 0.0971334 & 2486.60 & 1093.83  \\
10 & 0.0971334 & 2465.52 & 1098.30  \\
$\cdot$&$\cdot$ &$\cdot$& $\cdot$\\[-5pt]
$\cdot$&$\cdot$ &$\cdot$& $\cdot$\\[-5pt]
$\cdot$&$\cdot$ &$\cdot$& $\cdot$\\
15622&0.106343&3824.09 &2200.10 \\
15623&0.106343&3823.88 &2200.03 \\
15624&0.106343&3823.43 & 2199.99\\
15625&0.106343&3823.07 & 2199.95\\
15626&0.106343&3823.03 & 2199.91\\
15627&0.106343&3823.03 & 2199.87\\
15628&0.106343&3823.03 & 2199.84\\
15629&0.106343&3823.03 & 2199.82\\
15630&0.106343& 3823.03&2199.81 \\
15631&0.106343&3823.03 & 2199.81\\
\hline
\end{tabular}
\end{center}
\caption{\small{Layer number, layer thickness $h_i$ in~{\rm m}, and  corresponding $P$-wave and $S$-wave speeds in ${\rm m\,s}^{-1}$\,.}}
\label{table:well-log}
\end{table}
\section{Ray velocity in terms of ray angle}
\label{app:rayvelangle}
To derive the ray velocity from the ray angle in a transversely isotropic medium, we need equations (9.2.19),  (9.2.23), (8.4.9), and (8.4.12) of Slawinski (2015), which in our notation, using the density-scaled elasticity parameters, are 
\begin{equation}
\label{eq:qPwavevel}
v_{qP}(\vartheta)=\sqrt{\dfrac{\left(c^{\overline{\rm TI}}_{3333}-c^{\overline{\rm TI}}_{1111}\right)\cos^2\vartheta+c^{\overline{\rm TI}}_{1111}+c^{\overline{\rm TI}}_{2323}+\sqrt{\Delta}}{2}}
\,,
\end{equation}
where
\begin{align}
\Delta:=&\left(\left(c^{\overline{\rm TI}}_{1111}-c^{\overline{\rm TI}}_{2323}\right)\sin^2\vartheta-\left(c^{\overline{\rm TI}}_{3333}-c^{\overline{\rm TI}}_{2323}\right)\cos^2\vartheta\right)^{2} \nonumber \\
&+4\left(c^{\overline{\rm TI}}_{2323}+c^{\overline{\rm TI}}_{1133}\right)^{2}\sin^2\vartheta\cos^2\vartheta
\,,
\end{align}

\begin{equation}	
V\left(\vartheta\right)=\sqrt{\left[v\left(\vartheta\right)\right]^{2}+\left[\frac{\partial
v\left(\vartheta\right)}{\partial\vartheta}\right]^{2}}\,,
\label{V(Q)isv}
\end{equation}

\begin{equation}	
\tan\theta=\frac{\frac{p_{1}}{p_{3}}+\frac{1}{v}\frac{\partial
v}{\partial\vartheta}}{1-\frac{p_{1}}{p_{3}}\frac{1}{v}\frac{\partial
v}{\partial\vartheta}}=\frac{\tan\vartheta+\frac{1}{v}\frac{\partial
v}{\partial\vartheta}}{1-\frac{\tan\vartheta}{v}\frac{\partial
v}{\partial\vartheta}}\,.
\label{PhaseGroup}
\end{equation}

The procedure is, given ray angle $\theta$\,, to numerically solve equation~(\ref{PhaseGroup}) for wavefront normal angle $\vartheta$\,, and then to use equation~(\ref{V(Q)isv}) to solve for ray velocity $V$.   Also note that

\begin{equation*}
\dfrac{\partial v_{qP}}{\partial\vartheta}=\dfrac
{\dfrac{D}
{C}
+A}
{B}
\end{equation*}

\begin{equation*}
A=-2 (c^{\overline{\rm TI}}_{3333} -c^{\overline{\rm TI}}_{1111} ) \sin (\vartheta ) \cos (\vartheta)
\end{equation*}

\begin{equation*}
B=2 \sqrt{2} \sqrt{E
+(c^{\overline{\rm TI}}_{3333} -c^{\overline{\rm TI}}_{1111} ) \cos ^2(\vartheta
)+c^{\overline{\rm TI}}_{1111} +c^{\overline{\rm TI}}_{2323} }
\end{equation*}

\begin{equation*}
C=2
\sqrt{\left[(c^{\overline{\rm TI}}_{1111} -c^{\overline{\rm TI}}_{2323} ) \sin ^2(\vartheta
)-(c^{\overline{\rm TI}}_{3333} -c^{\overline{\rm TI}}_{2323} ) \cos ^2(\vartheta )\right]^2+4
(c^{\overline{\rm TI}}_{1133} +c^{\overline{\rm TI}}_{2323} )^2 \sin ^2(\vartheta ) \cos ^2(\vartheta
)}
\end{equation*}

\begin{align*}
D&=2 (2
(c^{\overline{\rm TI}}_{1111} -c^{\overline{\rm TI}}_{2323} ) \sin (\vartheta ) \cos (\vartheta )+2
(c^{\overline{\rm TI}}_{3333} -c^{\overline{\rm TI}}_{2323} ) \sin (\vartheta ) \cos (\vartheta ))
\left[(c^{\overline{\rm TI}}_{1111} -c^{\overline{\rm TI}}_{2323} ) \sin ^2(\vartheta
) \right. \\  
&\left. -(c^{\overline{\rm TI}}_{3333} -c^{\overline{\rm TI}}_{2323} ) \cos ^2(\vartheta )\right]+8
(c^{\overline{\rm TI}}_{1133} +c^{\overline{\rm TI}}_{2323} )^2 \sin (\vartheta ) \cos ^3(\vartheta )-8
(c^{\overline{\rm TI}}_{1133} +c^{\overline{\rm TI}}_{2323} )^2 \sin ^3(\vartheta ) \cos (\vartheta )
\end{align*}

\begin{equation*}
E=\sqrt{\left[(c^{\overline{\rm TI}}_{1111} -c^{\overline{\rm TI}}_{2323} ) \sin
^2(\vartheta )-(c^{\overline{\rm TI}}_{3333} -c^{\overline{\rm TI}}_{2323} ) \cos ^2(\vartheta
)\right]^2+4 (c^{\overline{\rm TI}}_{1133} +c^{\overline{\rm TI}}_{2323} )^2 \sin ^2(\vartheta ) \cos
^2(\vartheta )}
\end{equation*}
\end{document}